\documentclass[prd,twocolumn,showpacs,preprintnumbers,amsmath,amssymb,superscriptaddress,amsmath,aps,10pt,nofootinbib]{revtex4-1}

\makeatletter
\renewcommand*\env@matrix[1][c]{\hskip -\arraycolsep
  \let\@ifnextchar\new@ifnextchar
  \array{*\c@MaxMatrixCols #1}}
\makeatother

\usepackage[bottom]{footmisc}
\usepackage[utf8]{inputenc}
\usepackage{amsmath,array}
\usepackage{graphicx}  
\usepackage{hyperref}
\usepackage[all]{hypcap}  
\usepackage{dcolumn}                 
\usepackage{bm}                          
\usepackage{epstopdf}
\usepackage{soul}
\usepackage{footnote}
\usepackage[draft]{todonotes} 
\usepackage{gensymb}

\usepackage{textgreek}
\usepackage{siunitx}
  \DeclareSIUnit\gauss{G}
  \DeclareSIUnit\cps{cps}
  \DeclareSIUnit\year{yr}
\usepackage{xparse}
\NewDocumentCommand\angrange{O{} m m}{\SIrange[parse-numbers=false, #1]{\ang[parse-numbers=true]{#2}}{\ang[parse-numbers=true]{#3}}{}}

\usepackage[capitalise]{cleveref} 
\usepackage[caption=false]{subfig}

\newcommand{\dd}{\mathrm{d}}

\makeatletter
\renewcommand\paragraph{\@startsection{paragraph}{4}{\z@}%
	{1.5ex\@plus 0ex \@minus -.2ex}%
	{1.5ex \@plus .2ex}%
	{\normalfont\normalsize\textit}}
\makeatother

\begin{document}
 

\title{A novel detector system for KATRIN to search for keV-scale sterile neutrinos}

\newcommand\AffMaxPlanck{Max Planck Institute for Physics, Föhringer Ring 6, 80805 München, Germany}
\newcommand\AffKIT{Karlsruhe Institute of Technology, Hermann-von-Helmholtz-Platz 1, 76344 Eggenstein-Leopoldshafen, Germany}
\newcommand\AffCEA{Commissariat \`{a} l'\'energie atomique et aux \'energies alternatives, Centre de Saclay, DRF/IRFU, 91191 Gif-sur-Yvette, France}
\newcommand\AffTUM{Technische Universität München, Arcisstraße 21, 80333 München, Germany}
\newcommand\AffHeidelberg{University of Heidelberg, Seminarstraße 2, 69117 Heidelberg, Germany}
\newcommand\AffOakRidge{Oak Ridge National Laboratory, 1 Bethel Valley Road, Oak Ridge, 37831 Tennessee, USA}
\newcommand\AffIAS{Institute for Advance Study, Technische Universit\"at M\"unchen, James-Franck-Str. 1, 85748 Garching}
\newcommand\AffAPC{Astroparticule et Cosmologie, Universit\'e Paris Diderot, CNRS/IN2P3, CEA/IRFU, Observatoire de Paris, Sorbonne Paris Cit\'e, 75205 Paris Cedex 13, France}
\newcommand\AffHLL{Max-Planck-Institut Halbleiterlabor, Otto-Hahn-Ring 6, 81739 M\"unchen, Germany}
\newcommand\AffXGLab{XGLab srl, Bruker Nano Analytics, Via Conte Rosso 23, Milano, Italy}
\newcommand\AffPoli{Politecnico di Milano, Dipartimento di Elettronica, Informazione e Bioingegneria, Piazza Leonardo da Vinci, 32, 20133 Milano, Italy}

\author{Susanne Mertens}
\affiliation{\AffMaxPlanck{}}
\affiliation{\AffTUM{}}
\author{Antonio Alborini}
\affiliation{\AffXGLab{}}
\author{Konrad Altenmüller}
\affiliation{\AffTUM{}}
\affiliation{\AffCEA{}}
\author{Tobias Bode}
\affiliation{\AffMaxPlanck{}}
\author{Luca Bombelli}
\affiliation{\AffXGLab{}}
\author{Tim Brunst}
\affiliation{\AffMaxPlanck{}}
\author{Marco Carminati}
\affiliation{\AffPoli{}}
\author{David Fink}
\affiliation{\AffMaxPlanck{}}
\author{Carlo Fiorini}
\affiliation{\AffPoli{}}
\author{Thibaut Houdy}
\affiliation{\AffMaxPlanck{}}
\author{Anton Huber}
\affiliation{\AffKIT{}}
\author{Marc Korzeczek}
\affiliation{\AffKIT{}}
\author{Thierry Lasserre}
\affiliation{\AffCEA{}}
\affiliation{\AffTUM{}}
\affiliation{\AffIAS{}}
\affiliation{\AffAPC{}}
\author{Peter Lechner}
\affiliation{\AffHLL{}}
\author{Michele Manotti}
\affiliation{\AffXGLab{}}
\author{Ivan Peric}
\affiliation{\AffKIT{}}
\author{David C. Radford}
\affiliation{\AffOakRidge{}}
\author{Daniel Siegmann}
\affiliation{\AffMaxPlanck{}}
\affiliation{\AffTUM{}}
\author{Martin Slezák}
\affiliation{\AffMaxPlanck{}}
\author{Kathrin Valerius}
\affiliation{\AffKIT{}}
\author{Joachim Wolf}
\affiliation{\AffKIT{}}
\author{Sascha Wüstling}
\affiliation{\AffKIT{}}

\

\date{\today}

\begin{abstract}
Sterile neutrinos are a minimal extension of the Standard Model of Particle Physics. 
If their mass is in the kilo-electron-volt regime, they are viable dark matter candidates. One way to search for sterile neutrinos in a laboratory-based experiment is via tritium-beta decay, where the new neutrino mass eigenstate would manifest itself as a kink-like distortion of the $\beta$-decay spectrum. The objective of the TRISTAN project is to extend the KATRIN setup with a new multi-pixel silicon drift detector system to search for a keV-scale sterile neutrino signal. In this paper we describe the requirements of such a new detector, and present first characterization measurement results obtained with a 7-pixel prototype system. 
\end{abstract}

\maketitle

\section{Introduction}
Several theories beyond the Standard Model (SM) predict the existence of a new type of neutrino, a so-called sterile neutrino, which owes its name to the fact that it would not take part in any SM interaction. Right-handed partners for the known purely left-handed neutrinos are an example of such sterile neutrinos. While being a minimal extension of the SM, sterile neutrinos can tackle a large number of open questions. Very heavy ($\geq$ GeV) sterile neutrinos\footnote{As sterile neutrinos are singlets under all SM gauge transformations, they can possess a Majorana mass of arbitrary scale. Throughout this paper we refer to sterile neutrinos as neutrino mass eigenstates, which, strictly speaking, are not purely sterile, but can have an admixture of the active neutrino flavors.} provide a natural explanation for the light mass of the active neutrinos via the See Saw Mechanism \cite{PhysRevLett.44.912} and could shed light on the matter-antimatter asymmetry of the universe via Leptogenesis~\cite{1475-7516-2010-09-001}. Electron-volt-scale sterile neutrinos can solve a number of anomalies found in short-baseline oscillation experiments, such as the well known reactor anomaly~\cite{eVWhitePaper}. The focus of this paper is on sterile neutrinos in the kilo-electron-volt range, which are a viable candidate for dark matter~\cite{PhysRevLett.72.17, PhysRevLett.82.2832, PhysRevLett.110.061801}. 
One notable feature of this candidate is that it can act as effectively cold or warm dark matter depending on its production mechanism in the early universe~\cite{Zhu:2015, 1475-7516-2017-11-046}. 

The strongest experimental bounds on the existence of keV-scale sterile neutrino dark matter are currently obtained from astrophysical observations with telescopes, that search for a mono-energetic X-ray line arising from the decay of relic sterile neutrinos. These measurements limit the active-to-sterile mixing amplitude $\sin^2\Theta$ (which determines their interaction strength) to $\sin^2\Theta < 10^{-6} - 10^{-10}$ in a mass range of $1 - 50$ keV~\cite{1475-7516-2012-03-018, Boyarsky_XMMNewton}. Furthermore, cosmological considerations and observations of structure formation can indirectly limit the allowed parameter space for sterile neutrino dark matter; for a detailed review see \cite{1475-7516-2017-01-025, Boyarsky:2018tvu}. Current laboratory-based limits are orders of magnitudes weaker than astrophysical bounds~\cite{1475-7516-2017-01-025, Boyarsky:2018tvu}.  

\begin{figure}[]
	\centering
    \includegraphics[width=\columnwidth] {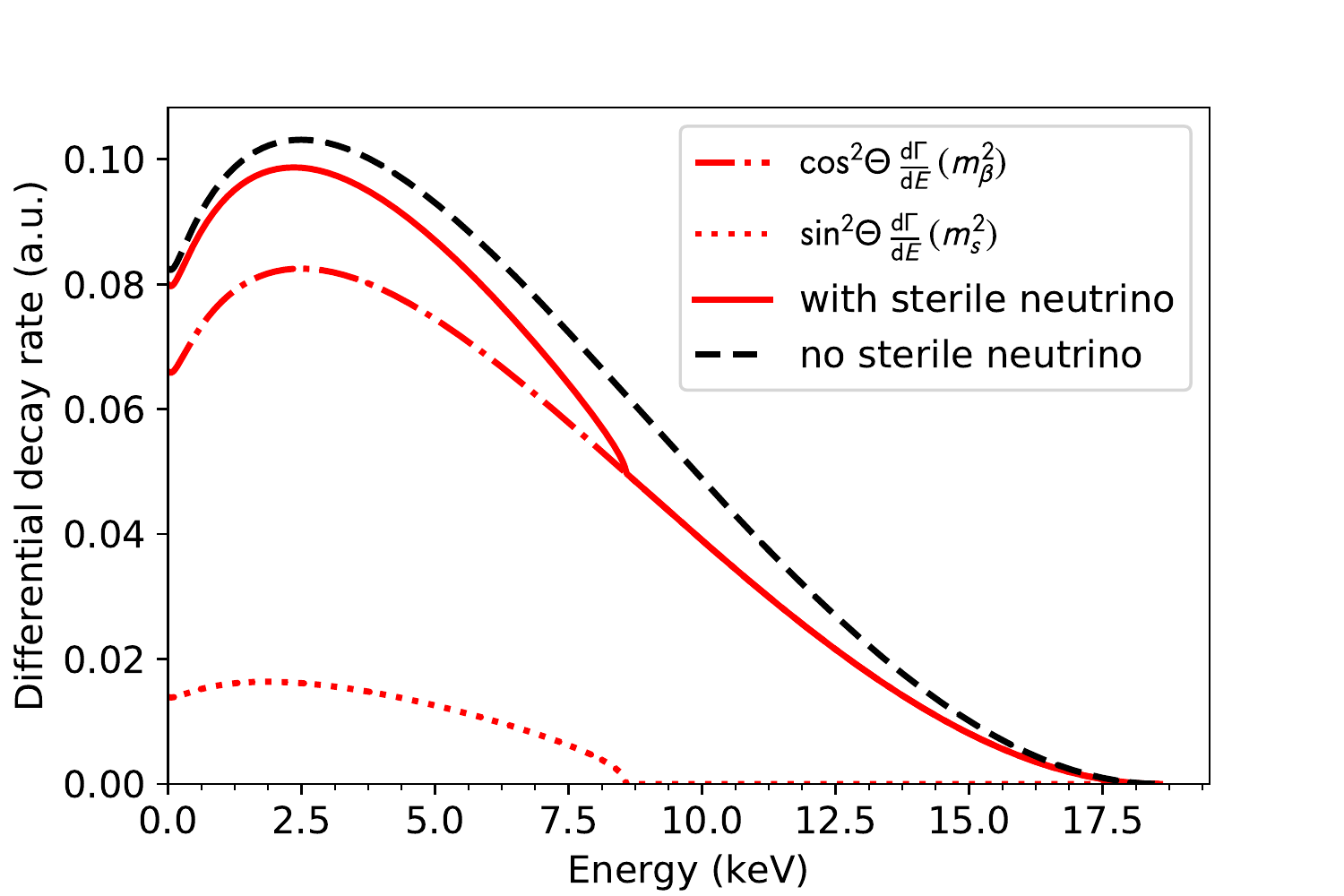}
    \label{fig:BetaSpectrum}
	\caption{Imprint of a heavy, mostly sterile, neutrino with a mass of $m_s = \SI{10}{\kilo\electronvolt}$ and an unphysical large mixing angle of $\sin^{2}\Theta=0.2$ on the tritium \textbeta{}-decay spectrum.}
\end{figure}

A unique way to perform a laboratory-based sterile neutrino search is via \textbeta{}-decay~\cite{SHROCK1980159, DEVEGA2013177, Mer:2015a, Mer:2015b}. In a \textbeta{}-decay an electron-flavor neutrino is emitted along with the electron. The corresponding continuous \textbeta{}-decay spectrum is a superposition of spectra corresponding to the different neutrino mass states $m_i$ that comprise the electron neutrino flavor eigenstate. Due to the tiny mass differences of the three light neutrino mass eigenstates, this superposition can not be resolved with current experiments such as the Karlsruhe Tritium Neutrino (KATRIN) experiment~\cite{Angrik:2005ep, Drex13}. However, sterile neutrinos with a mass $m_s$ significantly heavier than the known neutrino mass eigenstates $m_{1-3}$, would lead to a distinct distortion of the \textbeta{}-decay spectrum. The total differential tritium \textbeta{}-decay spectrum 
\begin{equation}
\frac{\dd\Gamma}{\dd E} = \cos^2\Theta \, \frac{\dd\Gamma}{\dd E} \left( m_\beta^2 \right) + \sin^2\Theta \, \frac{\dd\Gamma}{\dd E} \left( m_s^2 \right)
\end{equation}
would be a superposition of the standard spectrum, with the endpoint governed by the effective electron neutrino mass $m_\beta$, and a spectrum with a significantly reduced endpoint corresponding to the decay into a sterile neutrino of mass $m_s$. The amplitude of the additional decay branch would be governed by the active-sterile mixing amplitude $\sin^2\Theta$. Hence, sterile neutrinos would manifest themselves as a local kink-like feature and a broad spectral distortion below $E_\mathrm{kink}=E_0-m_\mathrm{s}$, see figure~\ref{fig:BetaSpectrum}. 

With an endpoint of $E_0 = 18.6$~keV the super-allowed \textbeta{}-decay of tritium is well suited for a keV-scale sterile neutrino search. Due to the short-half life of 12.3 years, high decay rates can be achieved with reasonable amounts of tritium. Furthermore, a kink-like sterile neutrino signature would be a distinct feature in the fully smooth tritium $\beta$ decay spectrum~\cite{Mer:2015a}. 

\section{The TRISTAN project}
\begin{figure*}[]
	\centering
    \includegraphics[width=2\columnwidth] {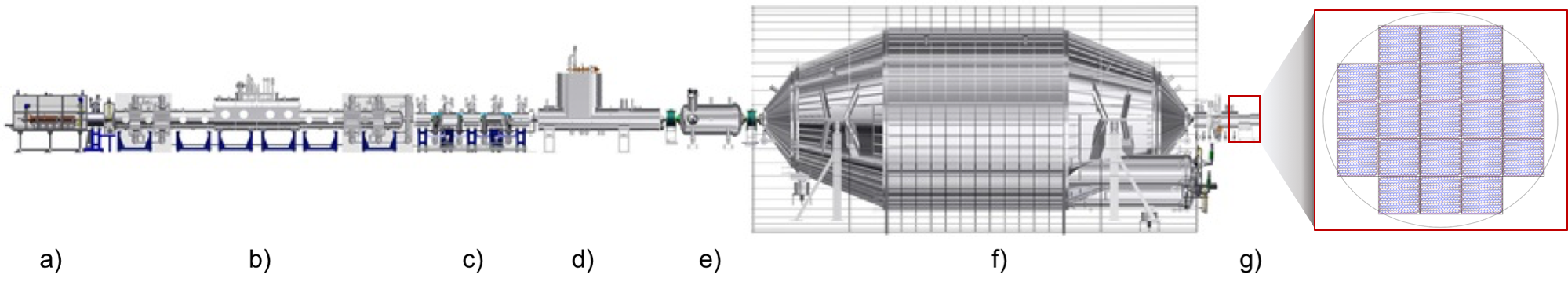}
    \label{fig:KATRINSetup}
	\caption{The left-hand side of the figure depicts the KATRIN apparatus: a) Rear section, b) windowless gaseous tritium source (WGTS), c) differential pumping section, d) cryogenic pumping section, e) pre-spectrometer, f) main spectrometer, g) detector section. For a detailed overview of KATRIN see reference~\cite{Angrik:2005ep}. For the TRISTAN project the 148-pixel Si-PIN focal plane detector~\cite{Amsbaugh201540} would be replaced by a $\approx$3500-pixel silicon drift detector system, displayed enlarged on the right-hand side of the figure. The baseline design comprises 21 so-called detector modules, each with 166 pixels and a side length of 4~cm. Each pixel has a hexagonal shape and a diameter of 3~mm. The total detector diameter is about 20~cm.}
\end{figure*}

\begin{figure}[]
\centering
		\includegraphics[width=\columnwidth]{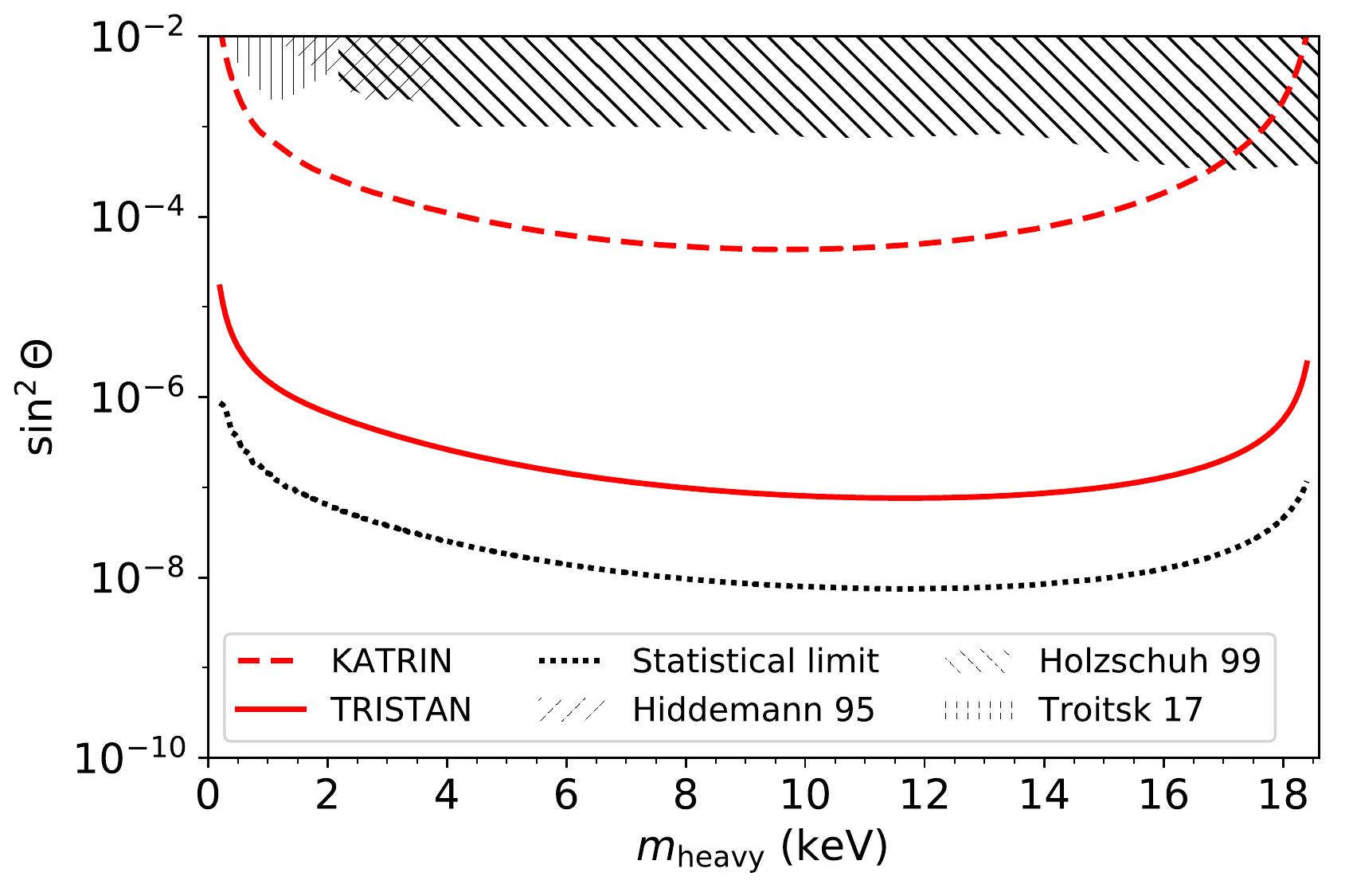}\\
		\label{fig:Sensitivity}
        \caption{This figure shows the 95\% C.L. sensitivity of 1) what could be achieved with KATRIN as it is at the moment (red dashed curve), 2) the design sensitivity of TRISTAN (red), and 3) the statistical limit that could be reached after a three years data-taking phase with the full source strength of the KATRIN experiment. For the TRISTAN measurement we assume a 3-years measurement period, with a 100 fold reduced tritium column density in order to reduce systematic effects related to scattering of electrons with gas molecules and pile-up effects. Furthermore, we assume a constant energy resolution of 300~eV FWHM. The grey dashed area depicts the current laboratory-based limit~\cite{Abdurashitov2017, Holzschuh:1999vy, 0954-3899-21-5-008}.}
\end{figure}

The idea of the TRISTAN project is to utilize the unprecedented tritium source luminosity of the KATRIN experiment for a high-precision keV-scale sterile neutrino search. TRISTAN is designed to achieve a sensitivity of $\sin^{2}\Theta<10^{-6}$, see figure~\ref{fig:Sensitivity}.

KATRIN's prime goal is a direct probe of the absolute neutrino mass scale via a precise measurement of the tritium beta decay spectrum close to its endpoint, where the imprint of the neutrino mass is maximal. For this purpose, KATRIN combines a high-activity ($10^{11}$ decays per second) gaseous tritium source with a high-resolution ($\Delta E \sim 1$\,eV) spectrometer, see figure~\ref{fig:KATRINSetup}. The electrons are guided along magnetic field lines from the so-called Windowless Gaseous Tritium Source (WGTS) to the spectrometer. The latter is operated as a Magnetic Adiabatic Collimation and Electrostatic (MAC-E) filter~\cite{0022-3735-16-4-016, PICARD1992345}, which transmits electrons with kinetic energies larger than the spectrometer's retarding potential. By observing the number of transmitted electrons for different filter voltages $U$ in a range of about $E_0 - 60~\text{eV} < qU < E_0 + 5~\text{eV}$ (where q is the electron charge) the integral tritium \textbeta-decay spectrum is obtained. The rate of electrons is detected with a 148-pixel focal plane detector~\cite{Amsbaugh201540, Wall201473} situated at the exit of the KATRIN spectrometer. The detector system is equipped with a post-acceleration electrode (PAE), that increases the kinetic energy of all electrons by a fixed amount of up to 20~keV~\cite{Amsbaugh201540}.

Operating KATRIN to search for keV-scale sterile neutrinos requires extending the measurement interval to cover the entire tritium \textbeta-decay spectrum, i.e.\ to set the filter voltage to values much lower than in standard operation~\cite{Mer:2015a}. In this new mode of operation the number of transmitted electrons will be a few orders of magnitudes higher than in normal KATRIN operation mode. The current silicon focal plane detector is not designed to handle such high count rates. Accordingly, the main objective of the TRISTAN project is to develop a new detector and read-out system, capable of revealing very small spectrum distortions, and handling rates of up to $10^{8}$ counts per second (cps).

The main challenge of a keV-scale sterile neutrino search is the precise understanding of the entire spectrum on the parts-per-million (ppm) level, in order to be able to start probing sterile-active mixing angles of cosmological interest. In order to reduce systematic uncertainties and avoid false-positive signals, a combination of an integral and a differential measurement mode is planned: The integral mode makes use of the high-resolution spectrometer and a counting detector (analogous to the normal KATRIN measurement mode). This mode requires an extremely stable counting rate. In the differential mode, the spectrometer is operated at low filter voltage continuously and the detector itself determines the energy of each electron. This mode requires an excellent energy resolution and a precise understanding of the detector response even at high counting rates. The two measurement modes are prone to different systematic uncertainties and hence allow to cross check each other. 

\begin{figure*}[]
	\subfloat[]{\includegraphics[width=0.545\columnwidth]{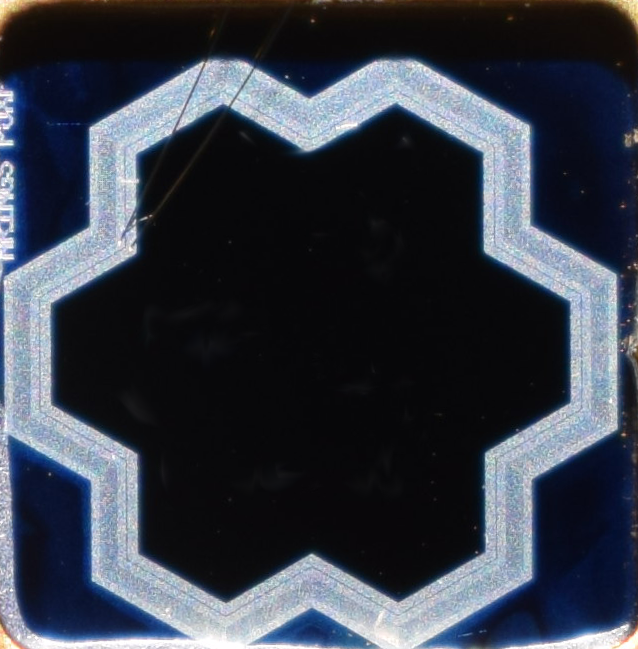}\label{fig:DetEntrance}} \hspace{0.5cm}
    \subfloat[]{\includegraphics[width=0.6\columnwidth]{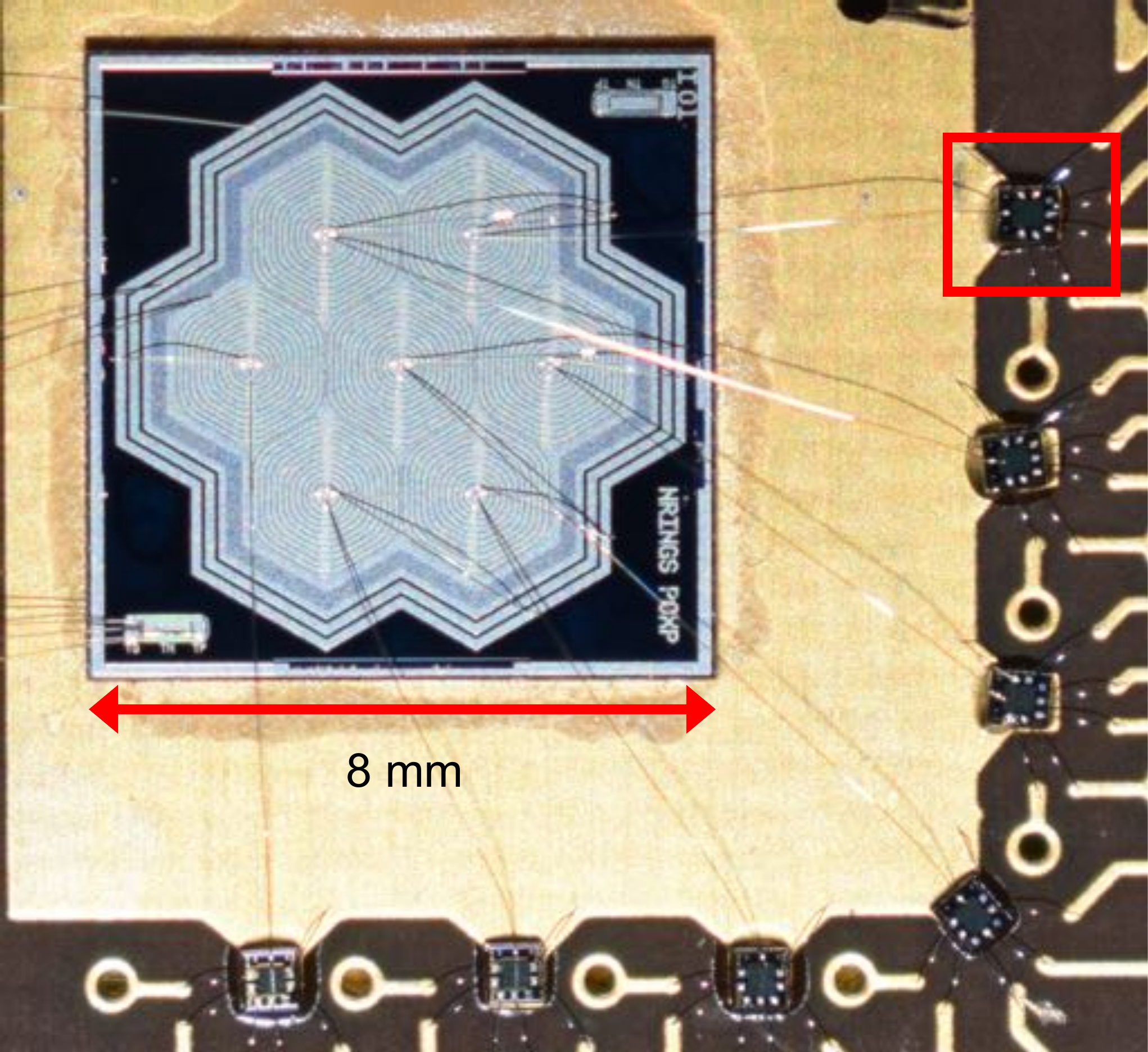}\label{fig:DetReadout}}
	\caption{Close-up photographs of the detector chip. \protect\subref{fig:DetEntrance} The photograph shows the entrance window side of the detector which has no structuring of the individual cells. A common electrode for depletion voltage and guard-ring surrounds the detector. \protect\subref{fig:DetReadout} Read-out side of detector chip. The drift rings are arranged around each individual cell anode and in this way define the cell shape and size. The anodes are bonded to individual CUBE preamplifier ASICs (one of them is circled in red square) which are arranged around the chip.}
\end{figure*}

TRISTAN is currently an R\&D effort for an experiment to take place after completion of the neutrino mass measurement of KATRIN, prospectively in 2025. In this paper we 1) discuss the requirements of the new detector and read-out system and 2) present the first characterization measurements of a 7-pixel prototype silicon drift detector, see figure~\ref{fig:DetEntrance}, produced at the Semiconductor Laboratory of the Max Planck Society (HLL)~\cite{HLL} and equipped with a read-out ASIC developed at the XGLab company~\cite{XGLab}.  

\section{Requirements on the detector system}
The current baseline design of the TRISTAN detector is a 3500-pixel Silicon Drift Detector (SDD) of about 20~cm diameter. SDDs are optimized for an excellent energy resolution (close to the Fano-limit for silicon) at high count rates and large-area coverage~\cite{Gatti:1984uu}, which makes them ideally suited for a keV-scale sterile neutrino search with KATRIN. In the following we detail the requirements and corresponding design choices with respect to pixel number, pixel size, energy resolution, and read-out electronics.  

\paragraph{Number of pixels}
The final TRISTAN detector is designed with the goal of at least achieving a sensitivity to active-sterile mixing angles in the $<$ ppm-level, see figure~\ref{fig:Sensitivity}. Correspondingly, the systematic uncertainties have to be on the same order of magnitude. From a purely statistical point of view, a ppm sensitivity can be reached with a total statistics of $10^{16}$ electrons. Assuming a data taking period of 3 years, this leads to a count rate of $n_{{tot}}=10^8$ cps on the entire detector. To minimize the pile-up probability a count rate per pixel of maximally 100~kcps is foreseen. This leads to a minimal pixel number of 1000 pixels. The total number of pixels is limited to keep the complexity and cost at a manageable level. This is especially true as a sophisticated front- and back-end readout electronics will be needed to ensure low noise, good energy resolution, high linearity, and the ability to tag charge sharing, and backscattering.  

\paragraph{Pixel and detector size}
In order to maximize statistics and avoid unused pixels, the detector size should be close to the size of the electron beam. The magnetic flux $\Phi = B \cdot A$, where $B$ is the magnetic field and $A$ is the beam area, is conserved along the entire KATRIN beamline. Hence, the electron beam area at the detector position is determined by the corresponding magnetic field $B_{\text{det}}$. This field can be adjusted rather easily, which in turn allows for variable detector diameters.

The optimization of the pixel and detector size is driven by the aim of minimizing both charge-sharing between neighboring detector pixels and backscattering of electrons from the detector surface. 

Minimal charge-sharing is obtained by choosing the largest pixel area that does not compromise the energy resolution and the charge collection time of the detector. The backscattering effect impacts the choice of pixel and detector size in multiple ways:
\begin{itemize}
\item To minimize the backscattering probability the pitch angle (angle between electron's momentum vector and magnetic field vector) at the detector should be $\theta_{\text{det}}<20^{\circ}$. The maximal pitch angle is determined by $\theta_{\text{max}} = \arcsin(\sqrt{\frac{B_{\text{det}}}{B_{\text{max}}}})$, and hence can be reduced by placing the detector in a magnetic field $B_{\text{det}}$ smaller than the maximal magnetic field in the system $B_{\text{max}} = 6~T$. The pitch angle is further reduced by applying a post-acceleration voltage $V_{\text{PAE}}$. For $V_{\text{PAE}} = 20~kV$, the desired pitch angle can be obtained with a detector magnetic field of $B_{\text{det}} \lesssim 1.5~T $. Correspondingly a minimal detector radius or $r_{\text{det}} \gtrsim 6$~cm is needed. 
\item By choosing $B_{\text{det}} < B_{\text{max}}$, backscattered electrons can be magnetically backreflected onto the detector surface. By increasing the backreflection probability the number of lost electrons can be reduced. The backreflection probability increases with the ratio $\frac{B_{\text{max}}}{B_{\text{det}}}$. This again supports the choice of a small detector magnetic field and consequently a large detector area.
\item Finally, it is preferred that the magnetically backreflected electrons hit the same or at most a neighboring pixel. In this case it would be possible to identify these events by applying a cut based on inter-arrival time for these spatially close events. The spatial distance of subsequent detector hits is mainly governed by the cyclotron radius of the electrons, which decreases with magnetic field as $r_{\text{cycl}}\propto\frac{1}{B}$. Consequently, to maximize the probability that a backreflected electron returns to the same or at most a neighboring pixel, the detector should be placed in a relatively large magnetic field, which would support the choice of a small detector area of $r_{\text{det}} \lesssim 11$~cm. 
\end{itemize}

A detailed optimization based on Monte-Carlo simulations was performed taking into account all above-mentioned effects~\cite{MarcThesis}. For this purpose the simulation software \textsc{KASSIOPEIA}~\cite{1367-2630-19-5-053012}, developed by the KATRIN collaboration, was used. It allows to compute the trajectory of electrons in the electromagnetic field setup of the KATRIN experiment. Moreover, it includes a module, called \textsc{KESS}~\cite{PascalThesis}, which simulates precisely the interactions of keV-scale electrons in silicon. 

In the simulation $\mathcal{O}(10^6$) electrons with energies up to 18~keV were started isotropically in the maximal magnetic field, and were tracked towards the detector, inside which interactions with silicon were simulated. Backscattered (primary and secondary) electrons are again tracked in the electromagnetic fields of the KATRIN vacuum system, where some of these electrons are reflected back to the detector by the maximum magnetic field. In this way, multiple backscatterings with the exact interarrival times and detector hit positions are taken into account. The simulations were performed for several detector positions, pixel sizes, and different post-acceleration voltages. Fixing the number of detector pixels to 3500 and the post-acceleration voltage to 20~keV, the optimal performance was found for a pixel diameter of 3~mm and a detector area of $A = \pi \cdot (10~\text{cm})^2$, corresponding to a detector magnetic field of 0.7~T. 

\begin{figure}[]
\centering
		\includegraphics[width=\columnwidth]{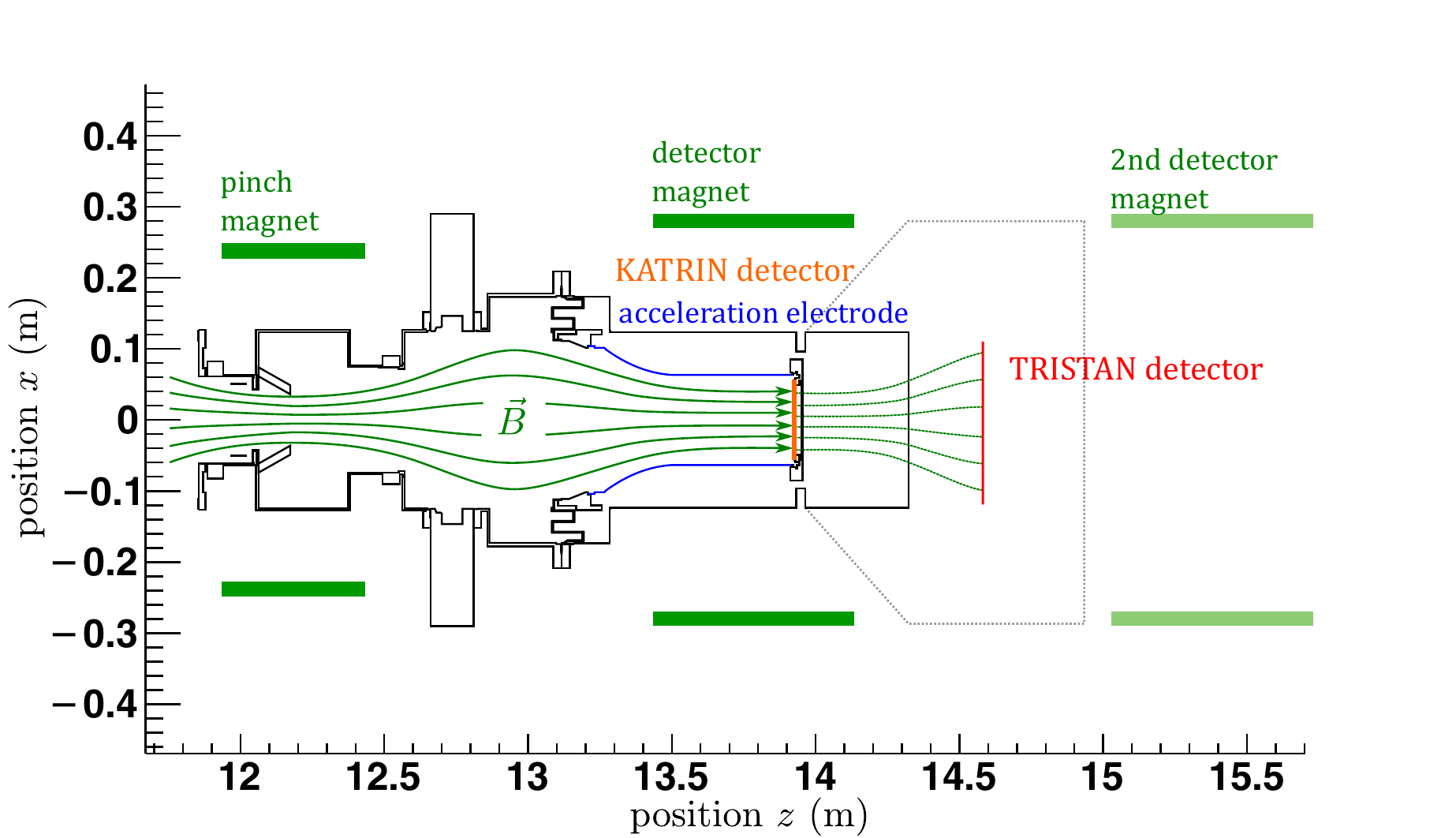}\\
		\label{Fig:Simulation}
        \caption{\textsc{Kassiopeia} simulation setup of the detector section of the KATRIN experiment. Electrons are guided by the magnetic field lines (green) onto the detector (red). The magnetic field is created by the pinch and detector magnet. An additional magnet can be placed behind the detector to make the field lines intersect perpendicularly with the detector surface. In blue the post-acceleration electrode (PAE) which can be set to up to 20~keV is displayed. In orange the current position of the KATRIN focal plane detector is shown.}
\end{figure}

\paragraph{Energy resolution}
To observe the kink-like structure an energy resolution of approximately 300\,eV at 30\,keV is needed ~\cite{Mer:2015b}. The SDD design allows for very low noise levels even for large pixel diameter (3~mm) and high count rates (100 kcps per pixel). Unlike in the case of X-ray detection, the challenge of the detection of electrons is that they unavoidably deposit energy in the non-active entrance window of the detector. For optimal energy resolution, an extremely thin entrance window of $\leq 100$~nm is needed. This poses a new challenge for the production technique of SDDs~\cite{FabricationUltraThin}. 

Charge sharing and backscattering can lead to partial small energy depositions. In order to be able to identify these events not only an excellent energy resolution but also a low energy threshold of about 1~keV is vital.

\paragraph{Read-out electronics}
A high performance read-out electronics is foreseen to assure high energy resolution and control of systematic effects.

To optimize the noise performance of the system, in the final system SDDs with integrated nJFET are planned to be used. In this case, the front-end ASIC only completes the feedback loop of the pre-amplification and can be placed at some distance from the detector anode. This is advantageous for a large-area detector, such as the TRISTAN detector array, where an installation of the pre-amplification close to the anode poses technical challenges. 

For the prototype system presented in this work, the SDD has no integrated nJFET. Correspondingly, an ASIC equipped with the FET had to be placed in close vicinity to the detector in order to minimize the input stray capacitance of the amplifier, which is a critical parameter in setting its noise performance.

A challenge of the back-end electronics is to minimize ADC non-linearities. ADC non-linearities can lead to discontinuous distortions of the differential \textbeta\,-decay spectrum, which could possibly mimic the signature of sterile neutrinos and hence reduce the sensitivity of the experiment. A detailed study demonstrated that using a waveform digitizing ADC (as opposed to a peak-sensing ADC) helps to significantly average out the effect of non-linearities~\cite{DOLDE2017127}. Moreover, waveform digitization with high sampling frequency of about 100 MHz helps to efficiently detect pile-up events. 

Finally, the field-programmable gate array (FPGA) of the back-end system has to provide the option of detecting not only pile-up events, but also multiplicities in neighboring or next-to-neighboring pixels. This option is needed to tag both charge-sharing and backscattering events. Fast processing and histogramming at the FPGA level is preferred for high-rate data taking phases in order to keep the data output at a manageable level. For characterization and validation, an event-by-event readout mode will also be necessary. 

\section{Prototype silicon-drift-detector setup}
A 7-pixel prototype SDD system was fabricated to test the general performance and to investigate the specific requirements mentioned above. Figure \ref{fig:DetEntrance} and \ref{fig:DetReadout} show photographs of the radiation entrance and read-out side of the sensor chip. 

\paragraph{Sensor chip}
As radiation sensor a custom made monolithic silicon drift detector array produced by the Semiconductor Laboratory of the Max Planck Society (HLL)~\cite{Lechner} was chosen. The novelty of these SDDs is that they feature a thin entrance window made possible by certain special processing steps of the p$^+$doped part of the pn-junction. For details on the sensor production see~\cite{FabricationUltraThin}.

Each of the seven pixels has an almost hexagonal shape and a characteristic small anode, 90~\textmu{}m, for a low detector capacitance of approximately 110~fF. The pixels are arranged in a gapless fashion so that no dead area inside the array is present. Arrays with a cell diameter of 0.25, 0.5, 1 and 2 mm were produced with a wafer thickness of 450~\textmu{}m. The number of drift rings was also varied. 

In this work we present results obtained with 1 and 2\,mm pixel diameter and 10 and 20 drift rings, respectively. The maximum pixel diameter of the prototype system is only 2~mm, even though a 3~mm pixel diameter is foreseen for the final system, since at the moment of the prototype production a slightly different design was foreseen.

\paragraph{Read-out chain}
The front-end read out electronics consists of a charge-sensitive preamplifier (CSA) application-specific integrated circuit (ASIC). The CUBE ASIC produced by the XGLab company~\cite{CUBE} was used, which achieves unprecedented noise performances, equivalent to SDD systems with integrated read-out~\cite{CUBE2}.

The CUBE ASIC is directly wire bonded to the pixel anode, see figure~\ref{fig:DetReadout}. The amplifier features a Field Effect Transistor (FET) in Complementary Metal Oxide Semiconductor (CMOS) technology whose high transconductance compensates possible additional capacitances by the wire bonds. The extremely low capacitance of the detector and front-end read-out leads to a low voltage (series) noise. This in turn allows operating the system at very short ($\leq$ 1~\textmu{}s) trapezoidal filter peaking times, and hence to achieve very good energy resolutions at high rates~\cite{CUBE2}. The CSA is a pulsed reset amplifier which is favorable for high rates since tail pile-up does not occur. Excellent energy resolution close to the Fano limit has been reported for this preamplifier, for details see \cite{CUBE}. 

As back-end read-out the DANTE digital pulse processor (DPP) also produced by XGLab was used. It is used to digitize the signal with a sampling frequency of 125 MS/s and a resolution of 16-bit. Furthermore, the DPP applies a trapezoidal filter with user-defined peaking and gap time for energy reconstruction.

\section{First characterization measurements}
The main objective of this first measurement phase was to test the general performance of the detector. In particular we focused on the energy resolution, linearity and noise performance of the system.

The measurements were performed at the Max Planck Institute for Physics in a dedicated setup, composed of a vacuum-tight stainless steel vessel with a cooling system, that can stably cool the detector to its optimal operating temperature of $-30^{\circ}$ Celcius. The setup features a special holder for a variety of radioactive calibration sources. All measurements presented here were performed with X-ray and gamma sources.

\paragraph{Energy resolution}
A suitable calibration source to test the performance of the system is $^{55}$Fe, which features two close-by X-ray lines (MnK$\alpha$ and MnK$\beta$) at 5.9~keV and 6.5~keV. 

\begin{figure}[]
\centering
		\includegraphics[width=\columnwidth]{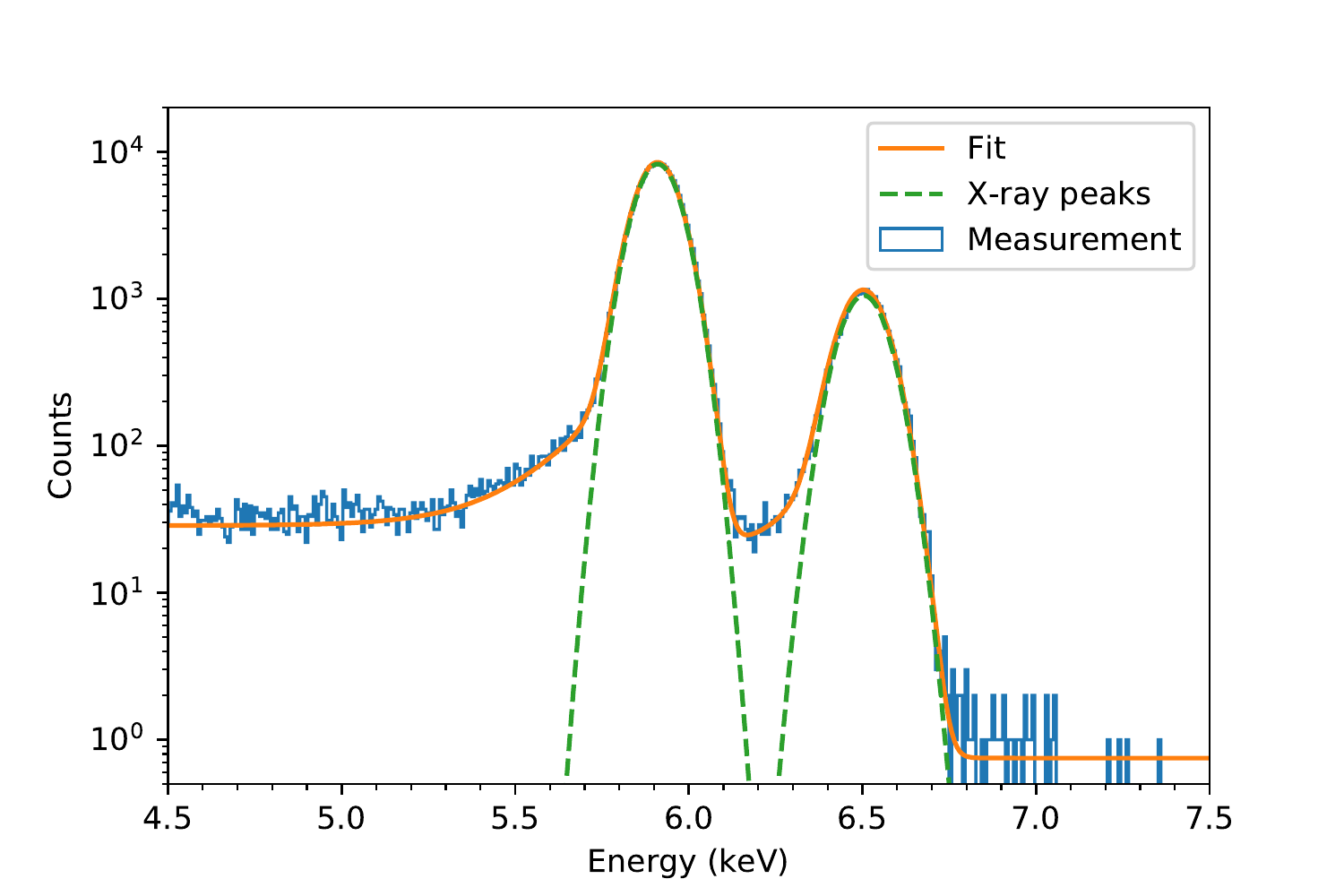}\\
		\label{Fig:Fe55}
        \caption{Energy spectrum of $^{55}$Fe recorded with a 2\,mm pixel diameter SDD array at -30\degree \,C. With an energy resolution of 139 eV (FWHM) 
@ 5.9 keV, the two lines are clearly separated. Two phenomenological functions similar to a hypermet (see \cite{Eggert}, pp. 102) were used to fit the spectrum (orange). The Gaussian parts of the two peaks are indicated in green.}
\end{figure}

A typical $^{55}$Fe spectrum recorded with a 2\,mm pixel-diameter SDD array at -30\degree \,C is shown in figure \ref{Fig:Fe55}. The main peaks can be well approximated by Gaussian functions and show an excellent energy resolution of 139~eV at the 5.9~keV x-ray line. 

The peak shapes show a low energy tail, which can be explained by incomplete charge collection close to the entrance window~\cite{Eggert, PeterModel}. This feature will be more pronounced for electrons, since, due to their shorter mean free path in silicon, they unavoidably deposit a fraction their energy in the non- or only partially-sensitive entrance volume. For this reason, for TRISTAN, SDDs with entrance windows of $\leq 100$~nm have been fabricated. Detailed characterizations with electron sources are currently ongoing.

\paragraph{Energy linearity}
To test the linearity of the system an $^{241}$Am calibration source was used. It emits photons in a wide energy from from 10 - 60~keV, which is precisely the energy region of interest for TRISTAN. The maximum energy reached in the experiment would be the tritium endpoint at 18.6~keV, shifted by the post-acceleration voltage of up to 30~keV. 

\begin{figure}[]
\centering
		\includegraphics[width=\columnwidth]{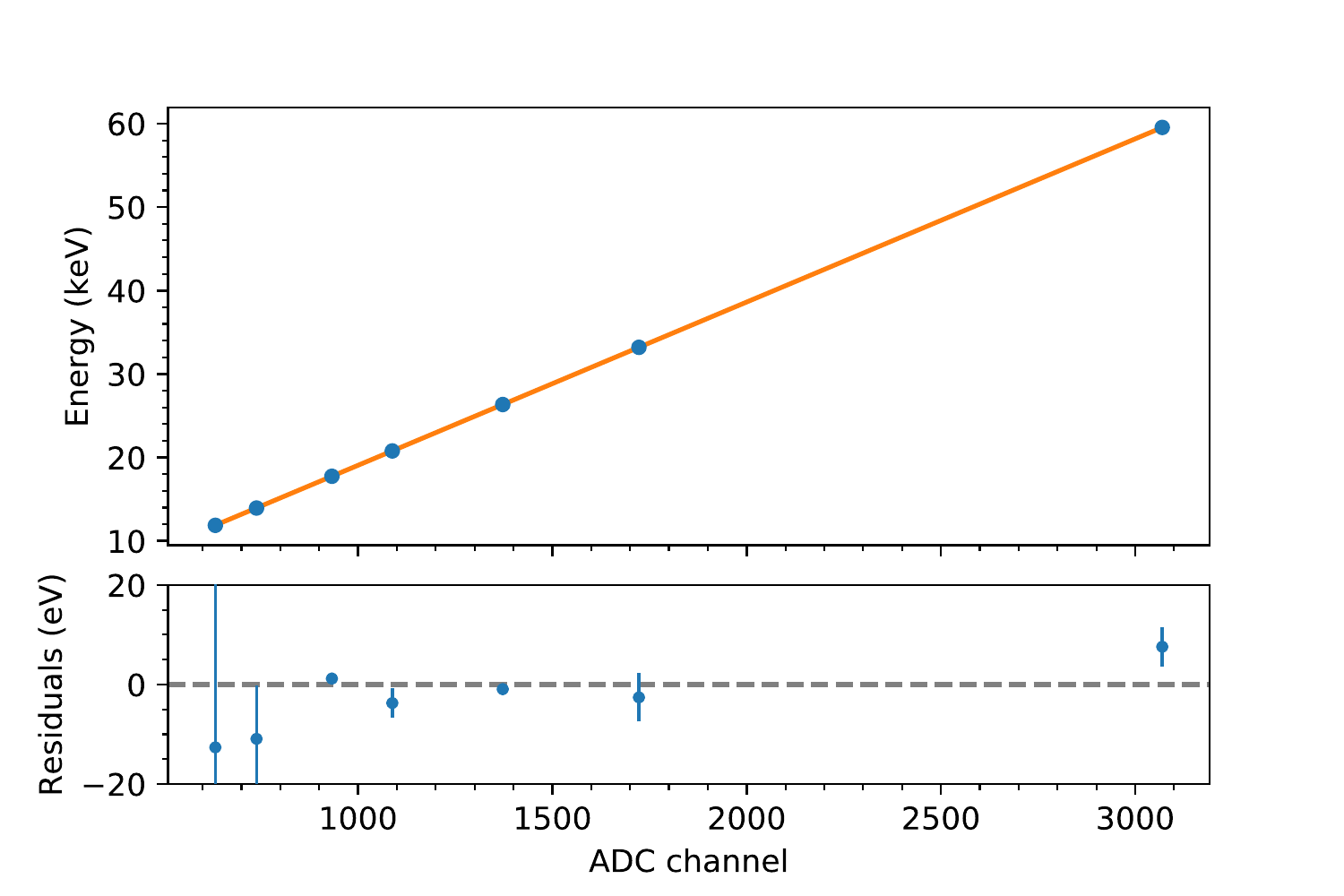}\\
		\label{Fig:Calibration}
        \caption{Linear calibration curve, based on an Am-241 spectrum. The blue dots correspond to the  following photon lines: 11.9~keV (x-ray), 13.9~keV (x-ray), 17.8~keV (x-ray), 20.8~keV (x-ray), 26.3~keV (gamma), 33.2~keV (gamma), and 59.5~keV (gamma). The error bars contain the uncertainty of the line positions and the fit uncertainty. The maximal deviation from the linearity is 0.1\% at the lowest energy peak of 11.9~keV.}
\end{figure}

The result, displayed in figure~\ref{Fig:Calibration}, shows an excellent linearity the entire energy range. All measured line positions deviate from linearity by $< 0.1\%$. For the final TRISTAN experiment ADC-non-linearities will play a major role, as discussed in~\cite{DOLDE2017127}, hence detailed investigation concerning the choice of the most suitable DAQ system are currently ongoing.

\paragraph{Noise performance}
To investigate the noise performance of the detector system, the energy resolution of the 5.9~keV MnK$\alpha$ line is measured as a function of the peaking time. The peaking time is a parameter of the trapezoidal filter, applied to deduce the amplitude and hence the energy of an event from the waveform. It represents the time span over which the waveform is averaged, and hence reflects which noise components are dominant in the signal.

\begin{figure}
\includegraphics[width=\columnwidth]{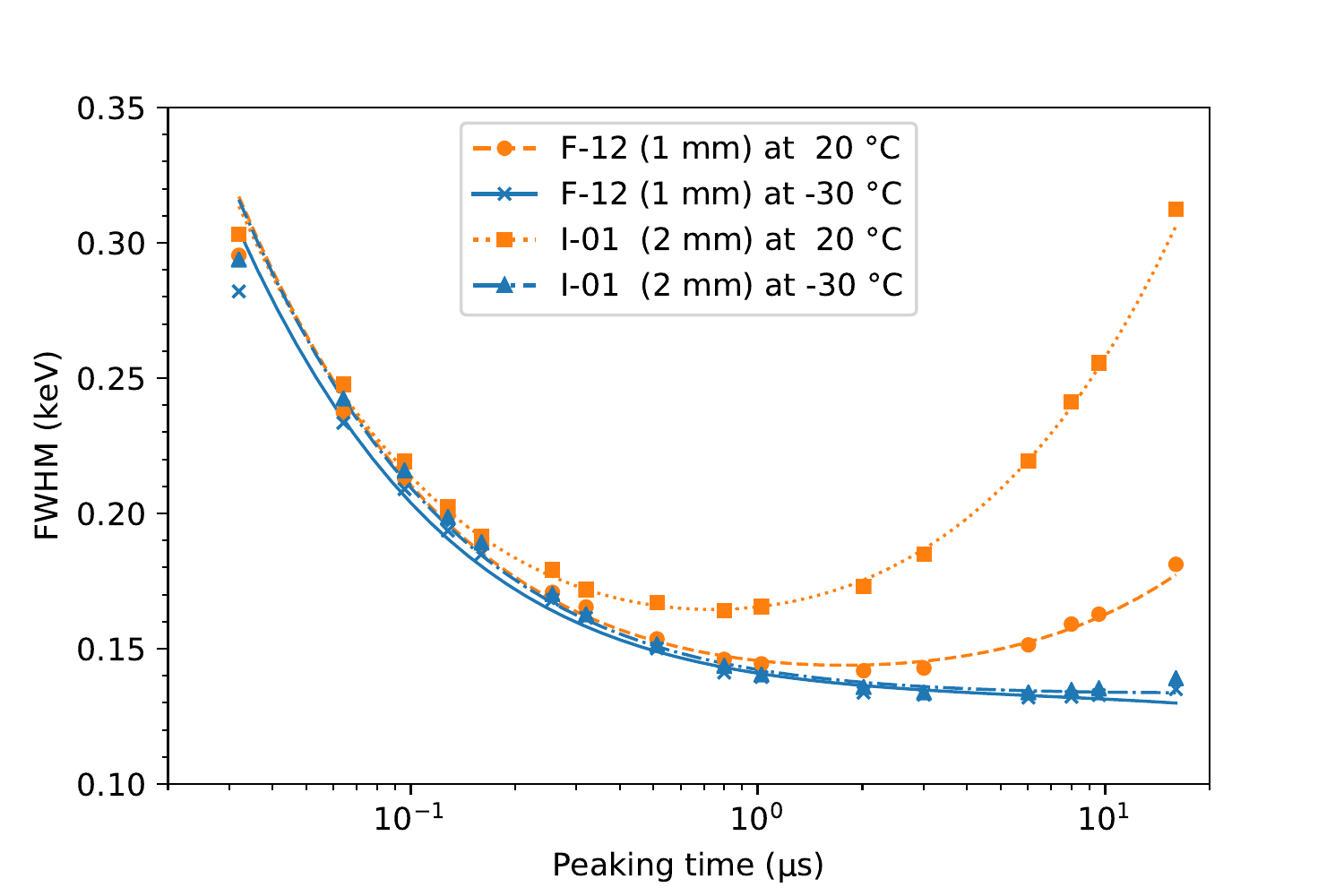}
        \label{Fig:FWHM_1vs2} 
        \caption{Noise curves for the 1-mm and 2-mm pixel-diameter detector at room temperature (orange) and $-30^{\circ}$~C (blue). At room temperature the rise of the noise for large peaking times is clearly visible and more pronounced for the larger pixel area, as expected. For cooled detectors the leakage current is reduced to a negligible level, such that the performance of both pixel dimensions is equivalent.}
\end{figure}

At room temperature the typical "U"-shaped noise curve is seen both for the 1~mm and 2~mm pixel diameter detector, see figure~\ref{Fig:FWHM_1vs2}. The increase of noise at large peaking times is due to leakage current. As expected, it is more pronounced for the 2~mm detector as the leakage current  grows with the area of the pixel. 

At $-30^{\circ}$ C, the leakage current is suppressed to a negligible level for both the 1~mm and 2~mm pixel diameter detector, see figure~\ref{Fig:FWHM_1vs2}. This demonstrates that the performance is not degraded for larger pixel areas. This is an important requirement for the TRISTAN detector, which will use a 3~mm pixel diameter.

Due to the lower leakage current and also reduced 1/f noise at $-30^{\circ}$~C, an energy resolution of FWHM of $\leq 140$~eV can be reached for peaking times of $\geq 1$~\textmu{}s. This corresponds to a minimal equivalent noise charge (ENC) value of $\leq 10$~e$_{\mathrm{rms}}$. 

It is important to note that even for peaking times of less than 1~\textmu{}s the energy resolution is still well below 300~eV (equivalently the ENC is $< 20$~e$_{\mathrm{rms}}$). As TRISTAN will operate at high rates of up to 100~kcps, a good energy resolution at short peaking times is essential. 

\section{Conclusion and Outlook}
A relevant search for a keV-mass sterile neutrino can be conducted with KATRIN experiment, providing that the current focal-plane detector is replaced with a new multi-pixel, high performance detector array, capable of coping with the high electron rates. 

Detailed simulations show that the stringent requirements of such an experiment can be met by a several-thousand-pixel SDD system. The SDD design provides a suitable energy resolution for large pixel areas and high rates. As a special request of the TRISTAN project, new techniques are being explored to minimize the entrance window thickness. 

A first seven-pixel prototype SDD array has been manufactured by the Semiconductor Laboratory of the Max Planck Society (HLL) and was equipped with preamplifier ASICs by XGLab S.R.L. First characterization measurements with the device verified the excellent performance of the  complete detector array, especially in terms of energy resolution at short peaking times. Measurements with dedicated electron calibration sources show first promising results, which will be reported in a separate paper.

A new 166-pixel sensor array is currently being produced by HLL, which will depict 1 of the 21 final TRISTAN detector modules. An integrated FET was chosen for the next prototype; the corresponding ASIC is in its prototyping phase. Both the 7-pixel SDD system and the new 166-pixel detector system will be used at the Troitsk nu-mass experiment~\cite{Troitsk} before being installed at the KATRIN setup, prospectively in 2025.

\section*{Acknowledgements}
We thank the Halbleiterlabor of the Max Planck Society, XGLab, and O. Limuousin from CEA for the fruitful cooperation. We gratefully acknowledge the support by the Max Planck Research Group (MPRG) program and especially the MaxPlanck@TUM initiative. 
Marc Korzeczek acknowledges the support by the DFG-funded Doctoral School "Karlsruhe School of Elementary and Astroparticle Physics: Science and Technology".
\bibliography{Paper1.bib}

\end{document}